\def \SAIT #1 #2 {{\em Mem.\ Soc.\ Astron.\ It.\/} {\bf #1}, #2}
\def \MESS #1 #2 {{\em The Messenger\/} {\bf #1}, #2}
\def \ASTRNACH #1 #2 {{\em Astron. Nach.\/} {\bf #1}, #2}
\def \AAP #1 #2 {{\em Astron. Astrophys.\/} {\bf #1}, #2}
\def \AAL #1 #2 {{\em Astron. Astrophys. Lett.\/} {\bf #1}, L#2}
\def \AAR #1 #2 {{\em Astron. Astrophys. Rev.\/} {\bf #1}, #2}
\def \AAS #1 #2 {{\em Astron. Astrophys. Suppl. Ser.\/} {\bf #1}, #2}
\def \AJ #1 #2 {{\em Astron. J.\/} {\bf #1}, #2}
\def \ANNREV #1 #2 {{\em Ann. Rev. Astron. Astrophys.\/} {\bf #1}, #2}
\def \APJ #1 #2 {{\em Astrophys. J.\/} {\bf #1}, #2}
\def \APJL #1 #2 {{\em Astrophys. J. Lett.\/} {\bf #1}, L#2}
\def \APJS #1 #2 {{\em Astrophys. J. Suppl.\/} {\bf #1}, #2}
\def \APSS #1 #2 {{\em Astrophys. Space Sci.\/} {\bf #1}, #2}
\def \ASR #1 #2 {{\em Adv. Space Res.\/} {\bf #1}, #2}
\def \BAIC #1 #2 {{\em Bull. Astron. Inst. Czechosl.\/} {\bf #1}, #2}
\def \JSQRT #1 #2 {{\em J. Quant. Spectrosc. Radiat. Transfer\/} {\bf #1}, #2}
\def \MN #1 #2 {{\em Mon. Not. R. Astr. Soc.\/} {\bf #1}, #2}
\def \MEM #1 #2 {{\em Mem. R. Astr. Soc.\/} {\bf #1}, #2}
\def \PLR #1 #2 {{\em Phys. Lett. Rev.\/} {\bf #1}, #2}
\def \PASJ #1 #2 {{\em Publ. Astron. Soc. Japan\/} {\bf #1}, #2}
\def \PASP #1 #2 {{\em Publ. Astr. Soc. Pacific\/} {\bf #1}, #2}
\def \NAT #1 #2 {{\em Nature\/} {\bf #1}, #2}

\documentstyle{memsait}
\input epsf.sty
\begin{opening}
\title{THE ASCA HARD SERENDIPITOUS SURVEY} 
\author{ROBERTO DELLA CECA$^1$, VALENTINA BRAITO$^2$, ILARIA CAGNONI$^3$, 
AND TOMMASO MACCACARO$^1$}
\institute
{$^1$Osservatorio Astronomico di Brera, Milan, Italy\\
 $^2$Osservatorio Astronomico di Padova, Padua, Italy\\
 $^3$SISSA, Trieste, Italy}
\date{} 
\end{opening}

\begin{document}

\oddpagefooter{}{}{} 
\evenpagefooter{}{}{} 
\ 
\bigskip

\begin{abstract}

We summarize here some of the main results we have reached so far in
the ASCA Hard Serendipitous Survey, a survey program conducted in
the 2-10 keV energy band. In particular we will discuss the number-flux
relationship, the 2-10 keV spectral properties of the sources, and the
cross-correlation of the ASCA sources with ROSAT sources.

\end{abstract}

\section{Introduction}

The ASCA Hard Serendipitous Survey (HSS) is a systematic search for
sources in the $2-10$ keV  band using data from the GIS2
instrument on board the ASCA Observatory.

\noindent
The specific aims of this project, initiated a few years ago at the 
{\it Osservatorio Astronomico di Brera}, are: 
a) to investigate the census of the X-ray sources shining in the hard
X-ray sky down to a flux limit of $\sim 10^{-13}$ erg cm$^{-2}$ s$^{-1}$,
b) to study their X-ray spectral properties,  
c) to test the Unification Model for AGN and 
d) to evaluate the contribution to the Cosmic X-ray Background (CXB)
from the different classes of X-ray sources.
Finally, this sample of 2-10 keV selected sources, along with other
complete samples  from ASCA and BeppoSAX survey programs 
(Ueda et al., 1999a,b; Giommi, Perri and Fiore, 2000) 
will be of fundamental importance in the next few years:
a) to fix the bright tail of the much deeper Chandra and XMM-Newton number
counts relationship and b) to provide sources bright enough for detailed 
X-ray follow up studies. 
 
\section{The ASCA HSS Sample}

The data considered for the ASCA HSS were extracted
from the public archive of 1629 ASCA fields (as of December 18, 1997).
The fields selection criteria, the data preparation and analysis,
the source detection and selection and the computation of the sky
coverage are described in detail in the ``pilot" paper 
(Cagnoni, Della Ceca and Maccacaro, 1998) and in Della Ceca et al. (1999a).
In summary, we have  considered 300 GIS2 images suitable for this
project (e.g. at high galactic latitude, not centered on bright or
extended targets, not centered on groups or associations of targets,
etc..), using only the events in the \mbox{2-10 keV} energy range and
within 20 arcmin from the detector center.  These images have been
searched for sources with a signal-to-noise (S/N) ratio greater than
4.0 (a more restrictive criterion than that adopted in the ``pilot"
paper of Cagnoni et al. (1998) where a S/N $\geq$ 3.5 was used).  A
sample of 189 serendipitous sources with fluxes in the range $\sim 1
\times 10^{-13} - \sim 8 \times 10^{-12}$ erg cm$^{-2}$ s$^{-1}$
has been defined.
The sky coverage is a function of the flux and it ranges from 
$\sim 0.8$ deg$^2$ for $f_x \geq 1 \times 10^{-13}$ erg cm$^{-2}$ s$^{-1}$, to 
$\sim 10$ deg$^2$ for $f_x \geq 1.75 \times 10^{-13}$ erg cm$^{-2}$ s$^{-1}$ 
and to 
$\sim 50$ deg$^2$ for $f_x \geq 4.3 \times 10^{-13}$ erg cm$^{-2}$ s$^{-1}$;
the total area of sky covered by the ASCA HSS is 
$\sim 71$ deg$^2$.   
Up to now 58 sources have been spectroscopically identified.  The
optical breakdown is the following: 1 star, 5 cluster of galaxies, 5 BL
Lacs, 40 Broad Line Type 1 AGN, and 7 Narrow Line Type 2 AGN.
However we stress that this sample of identified objects is
probably not representative of the whole population.
Full details on the ASCA HSS source sample will be reported in Della
Ceca at al. (2001, in preparation).

\section{The 2$-$10 keV LogN($>$S) - LogS}

In Figure 1 we summarize the results from different survey programs
obtained using ASCA and {\it Beppo}SAX data in the flux range
between
$\sim 1\times 10^{-14}$ erg cm$^{-2}$ s$^{-1}$ and
$\sim 1\times 10^{-11}$ erg cm$^{-2}$ s$^{-1}$. 
The {\it Chandra} extension of the LogN($>$S) - LogS  down to 
$\sim 1\times 10^{-15}$ erg cm$^{-2}$ s$^{-1}$, 
is discussed in Giacconi et al. (these proceedings).

\begin{figure}
\epsfxsize=11.5cm 
\vspace{-1.0cm}
\hspace{0.0cm}\epsfbox{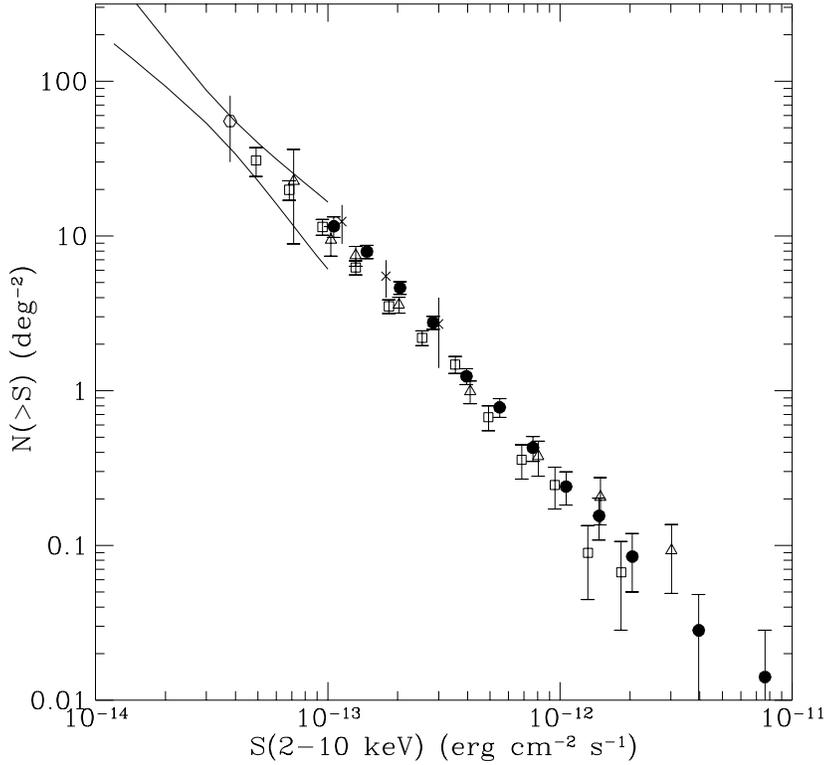}
\vspace{-0.5cm}
\caption[h]{The 2-10 keV LogN($>$S)-LogS 
from different surveys (adapted 
from Giommi, Perri and Fiore, 2000). 
Filled dot: ASCA HSS (this work); 
crosses: ASCA LSS (Ueda et al., 1999a);
open triangles: ASCA MSS (Ueda et al., 1999b);
open squares: {\it Beppo}SAX 2-10 keV survey (Giommi, Perri and Fiore, 2000);
open hexagon at $\sim 3.8\times 10^{-14}$ erg cm$^{-2}$ s$^{-1}$: ASCA DSS 
(Ogasaka et al., 1998);
solid line:  {\it Beppo}SAX  fluctuations analysis (Perri and Giommi, 2000).
}
\end{figure}

It can be seen that the different results are in good agreement  
and this is reassuring given the different instruments and/or data 
analysis utilized from the various groups. 

The ASCA HSS LogN($>$S)-LogS can be described by a power law model 
N($>$S) = $K \times S^{-\alpha}$ with best fit values ($\pm 1 \sigma$)  
of $\alpha = 1.67 \pm 0.09$ and 
$K=2.85^{0.168}_{46.2} \times 10^{-21}$ deg$^{-2}$
(note that K is not a fit parameter but is obtained by rescaling
the model to the actual number of objects in the sample).
The fit, performed over the flux range 
$\sim 8 \times 10^{-12}$ erg cm$^{-2}$ s$^{-1}$ $-$ 
$\sim 7\times 10^{-14}$ erg cm$^{-2}$ s$^{-1}$ 
(the faintest detectable flux), 
is in very good agreement with our previous computations 
(e.g. Cagnoni et al., 1998, Della Ceca et al., 1999b) 
and with the {\it Beppo}SAX results (Giommi, Perri and 
Fiore, 2000).
At the flux limit of the ASCA HSS survey ($\sim 7 \times
10^{-14}$ ergs cm$^{-2}$ s$^{-1}$) the total emissivity of the resolved
objects is  $\sim 4.5\times 10^{-12}$ erg cm$^{-2}$ s$^{-1}$ deg$^{-2}$, i.e.
about 20\% of the 2-10 keV CXB emissivity recently obtained 
from Vecchi et al. (1999) using {\it Beppo}SAX data. 

\section{The 2-10 keV spectral properties of the sources}

To investigate the spectral properties of the sources in the
2 $-$ 10 keV range we defined the hardness ratio, 
$HR2 = {H-M \over H+M}$, where M and H are the observed 
net counts in the 2$-$4 keV and 4$-$10 keV energy band
respectively (see Della Ceca et al.,1999a for details).
\begin{figure}
\epsfxsize=7.0cm 
\vspace{-0.6cm}
\hspace{-0.5cm}\epsfbox{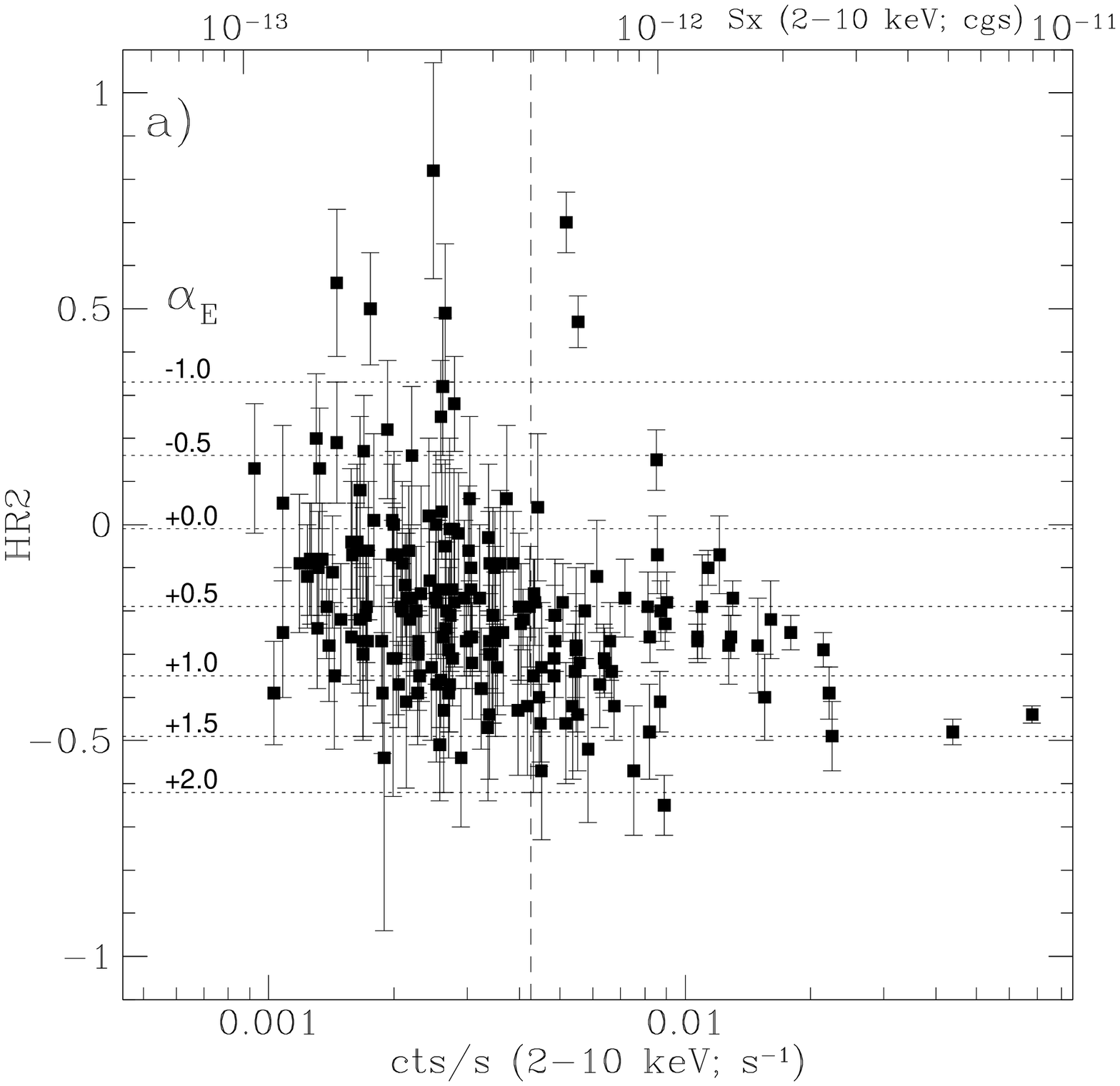}
\vspace{-7.0cm}
\epsfxsize=7.0cm
\hspace*{7.0cm}\epsfbox{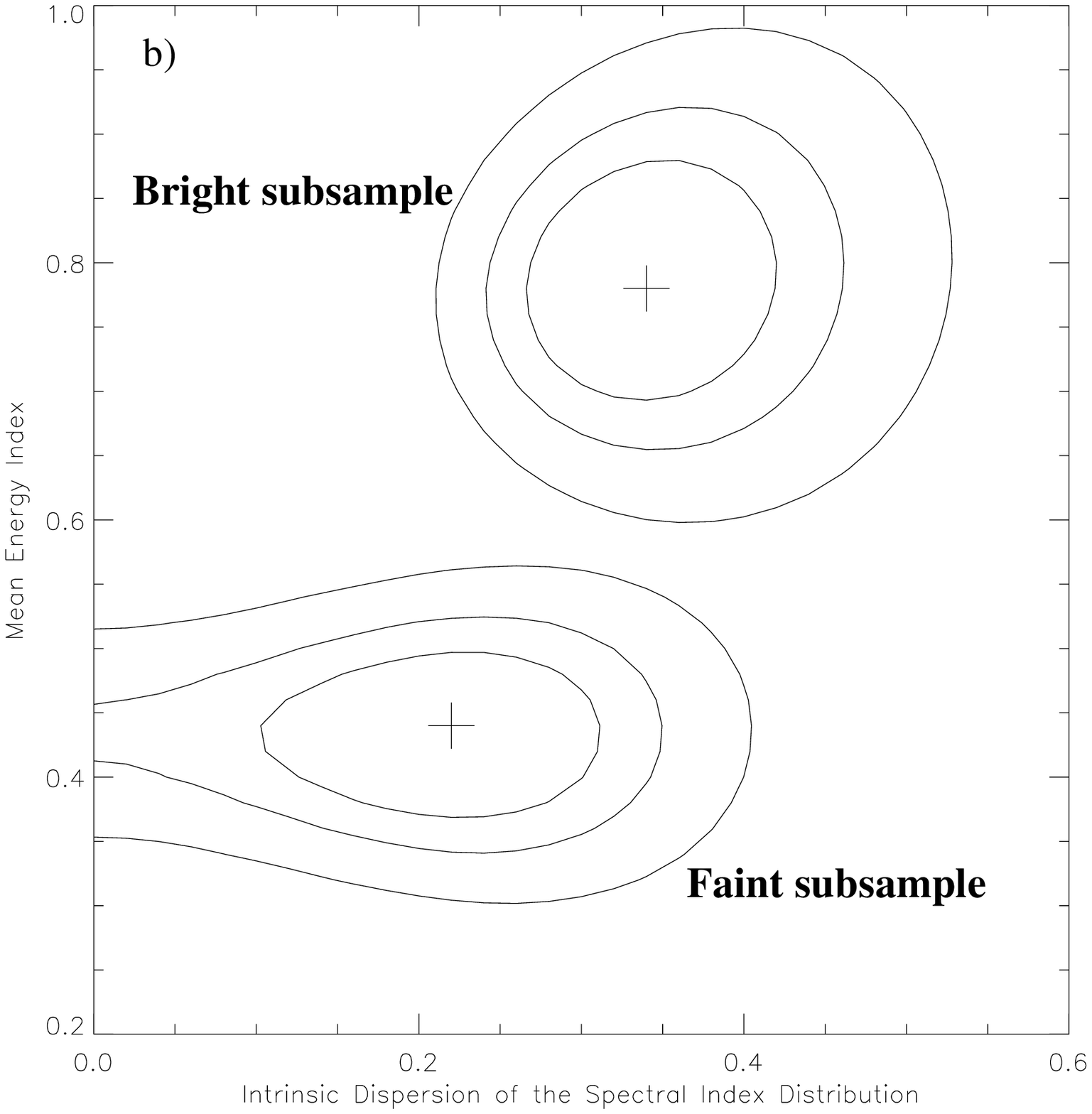} 
\vspace{-0.6cm}
\caption[h]{Panel a: HR2 vs. GIS2 count rate for the complete ASCA HSS
sample, compared with the HR2 expected from a non absorbed power-law
model with $\alpha_E$ ranging from $-$1.0 to 2.0.  The flux scale on
the top has been obtained assuming a conversion
factor appropriate for $\alpha_E \sim 0.6$, the median energy spectral
index of the sample.
Note the presence of many sources which seem
to be characterized by a very flat 2-10 keV spectrum with $\alpha_E
\leq 0.4$ and of a number of sources with ``inverted" spectra (i.e.
$\alpha_E \leq 0.0$).
Panel b: Confidence contours (68\%, 90\% and 99\% confidence level) of
the intrinsic dispersion of the spectral index distribution vs.  the
intrinsic mean energy index of the faint and bright subsets.  The
bright sample is defined by the 54 sources with a count rate $\geq
4.3\times 10^{-3}$ cts s$^{-1}$, while the faint sample is defined by
the remaining 128 sources. Seven sources with extreme spectral
properties ($\alpha_E < -1$  or $\alpha_E > 2$) have been excluded from
the analysis.
}
\end{figure}
Figure 2a (HR2 vs. GIS2 count rate) shows a flattening of the mean
spectrum of the sources with decreasing count rate as well as a large
spread in the HR2 (or $\alpha_E$) distribution; this spread is the
result of the convolution of the parent distribution with the
measurement error distribution.  To disentangle these two distributions
we have applied the maximum likelihood method as explained in
Maccacaro et al., 1988.  The results on the ``faint"  and on the ``bright"
subsets are reported in figure 2b.  We clearly detect a flattening of
the intrinsic  mean spectral properties of the sources which is significant 
at more than 99\% confidence level. 
Our results on the mean energy index for the faint subset are in good 
agreement with those obtained from Ueda et al., 1999a in a similar 
flux range  ($\alpha_E = 1.49 \pm 0.1$), while the results on the 
bright sample are in good agreement with the Ginga results 
(see Della Ceca et al., 1999a and reference therein).

\section{The ASCA-ROSAT cross correlation}

We  have cross-correlated the ASCA HSS sources with existing ROSAT
images (PSPC, HRI) and with the RASS catalogs.  
Only 11 of the 129 ASCA HSS
sources, for which we have adequate ROSAT data, do not have
a clear soft X-ray counterpart (all but 2 with $\alpha_E < 0.5$);
therefore we do not find a compelling evidence for a large ($> 9\%$)
population of 2-10 keV sources undetectable below a few keV (see also
Giommi, Perri and Fiore, 2000 for a similar results from the {\it
Beppo}SAX  2-10 keV survey).  In the popular and largely accepted
absorbed AGN scenario, this result implies that soft spectral
components must be present even in strongly absorbed objects as already
stated  from Giommi, Fiore and Perri, 1999 and Della Ceca et al.,
1999a.

\acknowledgements
\vskip -0.25cm
This work received partial financial support from MURST (Cofin-98-02-32). 
We thank P. Giommi and M. Perri
for helping us on the computation of the errors 
in the integral representation of the LogN-LogS and we thank   
A. Wolter having been able to retrieve, after 13 years, the 
ML code for the 
compution of the mean spectral properties. 


\end{document}